\documentclass[preprint,12pt]{aastex}
\renewcommand\>{{\rangle}}
\newcommand{\bld}[1]{\mbox{\boldmath $#1$}}

\newcommand{\dv}[2]{\frac{d#1}{d#2}}
\newcommand\ex{\hat{\bld{x}}}
\newcommand\ey{\hat{\bld{y}}}
\newcommand\ez{\hat{\bld{z}}}

\newcommand\<{{\langle}}

\shortauthors{Gammie \& Johnson}
\shorttitle{Vortices}
\begin{document}

\title{Vortices in Thin, Compressible, Unmagnetized Disks}

\author{Bryan M. Johnson and Charles F. Gammie}

\affil{Center for Theoretical Astrophysics,
University of Illinois at Urbana-Champaign,
1110 West Green St., Urbana, IL 61801}

\begin{abstract}

We consider the formation and evolution of vortices in a hydrodynamic
shearing-sheet model.  The evolution is done numerically using a version
of the ZEUS code.  Consistent with earlier results, an injected vorticity
field evolves into a set of long-lived vortices each of which has radial
extent comparable to the local scale height.  But we also find that the
resulting velocity field has positive shear stress $\<\Sigma \delta v_r
\delta v_\phi\>$.  This effect appears only at high resolution.  The
transport, which decays with time as $t^{-1/2}$, arises primarily because
the vortices drive compressive
motions.  This result suggests a possible mechanism for angular momentum
transport in low-ionization disks, with two important caveats: a
mechanism must be found to inject vorticity into the disk, and the
vortices must not decay rapidly due to three-dimensional instabilities.

\end{abstract}


\section{Introduction}

Astrophysical disks are common because the specific angular momentum of
the matter inside them is well-conserved.  They evolve because angular
momentum conservation is weakly compromised, either because of diffusion
of angular momentum within the disk or because of direct application of
external torques.  

In astrophysical disks composed of a well-ionized plasma it is likely
that some, perhaps most, of the evolution is driven by diffusion of
angular momentum within the disk.  This view is certainly consistent
with observations of steadily accreting cataclysmic variable systems
like UX Ursa Majoris \citep{bap98,bap04}, whose radial
surface-brightness profile is consistent with steady accretion-flow
models in which the bulk of the accretion energy is dissipated within
the disk.

Angular momentum diffusion in well-ionized disks is likely driven by
magnetohydrodynamic (MHD) turbulence.  Analytic analyses, numerical
experiments, and laboratory evidence strongly suggest that
well-coupled plasmas in differentially-rotating flows are subject to 
the magnetorotational instability (MRI; \citealt{bh91,bh98,bal03}).  
But MHD turbulence is initiated by the MRI only so long as the 
plasma is sufficiently ionized to couple to the magnetic field 
\citep{kb04,des04}.  In disks around young stars, cataclysmic-variable 
and X-ray binary disks in quiescence, and possibly the outer parts of AGN 
disks, the plasma may be too neutral to support magnetic activity
\citep{gm98,men00,sgbh00,mq01,ftb02}. This motivates interest in
non-MHD angular momentum transport mechanisms.

Within the last few years, a body of work has been developed suggesting
that vortices can be generated as a result of global hydrodynamic
instability \citep{haw87,bh88,haw90,love99,li00} or local hydrodynamic
instability \citep{kb03}, that vortices in disks may be long-lived
\citep{gl99,gl00,ur04,bm05}, and that these vortices may be
related to an outward flux of angular momentum \citep{li01,bm05}.
If these claims can be verified then the consequences for low-ionization
disks would be profound.

Here we investigate the evolution of a disk that is given a large
initial vortical velocity perturbation.  Our study is done in the
context of a (two-dimensional) shearing-sheet model, which permits us to
resolve the dynamics to a degree that is not currently possible in a
global disk model.  Our model is also fully compressible, unlike previous
work using a local model \citep{ur04,bm05}.  The former assume incompressible
flow and the latter use the anelastic approximation (e.g., \citealt{gou69}),
which filters out the high-frequency acoustic waves.  We will show that
compressibility and acoustic waves play an essential part in the angular
momentum transport.

Our paper is organized as follows.  In \S2 we describe the model.  In
\S3 we describe the evolution of a fiducial, high-resolution model.
In \S4 we investigate the dependence of the results on model
parameters.  And in \S5 we describe implications, with an emphasis on
key open questions: are the vortices destroyed by three-dimensional
instabilities?; and do mechanisms exist that can inject vorticity into the
disk?

\section{Model}

The shearing-sheet model is obtained via a rigorous expansion of the
two-dimensional hydrodynamic equations of motion to lowest order in
$H/R$, where $H = c_s/\Omega$ is the disk scale height ($c_s$ is the
isothermal sound speed and $\Omega$ is the local rotation frequency)
and $R$ is the local radius.  See \cite{ngg87} for a description. Adopting
a local Cartesian coordinate system where the $x$ axis is oriented
parallel to the radius vector and the $y$ axis points forward in
azimuth, the equations of motion become
\begin{equation}\label{EQ1}
\dv{\Sigma}{t} + \Sigma \bld{\nabla} \cdot \bld{v} = 0,
\end{equation}
\begin{equation}\label{EQ2}
\dv{\bld{v}}{t}
+ {\bld{\nabla} P\over{\Sigma}} + 2\bld{\Omega}\times\bld{v} - 2q\Omega^2 x
\ex = 0,
\end{equation}
where $\Sigma$ and $P$ are the two-dimensional density and pressure,
$\bld{v}$ is the fluid velocity and $d/dt$ is the
Lagrangian derivative. The third and fourth terms in equation
(\ref{EQ2}) represent the Coriolis and centrifugal forces in the local
model expansion, where $q = -(1/2) \, d\ln\Omega^2/d\ln r$ is 
the shear parameter.  We will assume throughout that $q = 3/2$, 
corresponding to a Keplerian shear profile.  We
close the above equations with an isothermal equation of state
\begin{equation}\label{EQ3}
P = c_s^2 \Sigma,
\end{equation}
where $c_s$ is constant in time and space.

Equations (\ref{EQ1}) through (\ref{EQ3}) can be combined to show
that the vertical component of potential vorticity
\begin{equation}\label{PVEV}
\xi \equiv \frac{(\bld{\nabla} \times \bld{v} + 2\bld{\Omega})\cdot \ez}{\Sigma}
\end{equation}
is a constant of the motion; i.e., the potential vorticity of fluid elements
in two dimensions is conserved.

An equilibrium solution to the equations of motion is
\begin{equation}
\Sigma = \Sigma_0 = const.
\end{equation}
\begin{equation}
P = c_s^2 \Sigma_0 = const.
\end{equation}
\begin{equation}
v_x = 0
\end{equation}
\begin{equation}
v_y = -q \Omega x
\end{equation}
Thus the differential rotation of the disk makes an appearance in the
form of a linear shear.

We integrate the above equations using a version of the ZEUS code
\citep{sn92}.  ZEUS is a time-explicit, operator-split scheme on a
staggered mesh.  It uses artificial viscosity to capture shocks.  Our
computational domain is a rectangle of size $L_x \times L_y$ containing
$N_x \times N_y$ grid cells.  The numerical resolution is therefore $\Delta x
\times \Delta y = L_x/N_x \times L_y/N_y$.

Our code differs from the standard ZEUS algorithm in two respects.
First, we have implemented a version of the shearing-box boundary
conditions.  The model is then periodic in the $y$ direction;  the $x$
boundaries are initial joined in a periodic fashion, but they are
allowed to shear with respect to each other, becoming periodic again
when $t = n L_y/(q\Omega L_x)$, $n = 1,2,\ldots$.  A detailed
description of the boundary conditions is given in \cite{hgb95}.

Second, we treat advection by the mean flow $\bld{v}_0 = -q\Omega x \ey$
separately from advection by the perturbed flow $\delta \bld{v} \equiv
\bld{v} - \bld{v}_0$.  Mean-flow advection can be done by interpolation, using
the algorithm described in \cite{gam01}, which is similar to the FARGO
scheme \citep{mass00}.  This has the advantage that the timestep is not
limited by the mean flow velocity (it is $|\delta \bld{v}|$ rather than
$|\bld{v}|$ that enters the Courant condition).  This permits the use of a
timestep that is larger than the usual timestep by $\sim L_x/H$ if $L_x
\gg H$.  The shear-interpolation scheme also makes the algorithm more
nearly translation-invariant in the $x-y$ plane, thereby more
nearly embodying an important symmetry of the underlying equations.

\subsection{Initial Conditions}

Without a specific model for the process that is injecting the
vorticity, it is difficult to settle on a particular set of initial
conditions, or to know how these initial conditions ought to vary
when the size of the box is allowed to vary.  Our choice of initial
conditions is therefore somewhat arbitrary. We use a set of initial
(incompressive) velocity perturbations drawn from a Gaussian random
field.  The amplitude of the perturbations is characterized by $\sigma =
\<|\delta \bld{v}/c_s|^2\>^{1/2}$.  The power spectrum is $|\delta \bld{v}|^2
\sim k^{-8/3}$, corresponding to the energy spectrum ($E_k \sim k^{-5/3}$)
of a two-dimensional Kolmogorov inverse turbulent cascade, with
cutoffs at $k_{min} = (1/2) (2\pi/H)$ and $k_{max} = 32 k_{min}$\footnote{We 
have compared our fiducial run to runs with a different range in $k$,
corresponding to vorticity injection either at scales $\sim H$ or scales
$\sim 0.1 H$.  The results are qualitatively the same.}.
The surface density is not perturbed. These initial conditions correspond
to a set of purely vortical perturbations.  The parameters for our fiducial
run are $L_x = L_y = 4H$ and $\sigma = 0.4$.

\subsection{Code Verification}

Although our basic algorithm has already been tested (see \citealt{gam01}),
we test the current version of our code by making a comparison with linear
theory.  Due to the underlying shear, small-amplitude perturbations in the
shearing sheet are naturally decomposed in terms of shearing waves
or {\it shwaves} (see \citealt{jg05} and references therein), Fourier
components in the ``co-shearing'' frame. These have time-dependent
wavenumber $\bld{k}(t) = k_x(t)\ex + k_y\ey$, where $k_x(t) = k_{x0}
+ q\Omega k_y t$ and $k_{x0}$ and $k_y$ are constant. The evolution of
a single Fourier component can be calculated by integrating an ordinary
differential equation for the amplitude of the shwave.  For purely vortical
(nonzero {\it potential} vorticity)
or non-vortical perturbations, the evolution can be obtained analytically. The
explicit expression for the amplitude of a vortical (incompressive) shwave is
\begin{equation}\label{IVX}
\delta v_{xi} = \delta v_{x0} \frac{k_0^2}{k^2} =
\delta v_{x0} \frac{1 + \tau_0^2}{1+\tau^2},
\end{equation}
where $k^2 = k_x^2 + k_y^2$, $\tau = q\Omega t + k_{x0}/k_y$ and a
subscript $0$ on a quantity indicates its value at $t = 0$.\footnote{This
solution is valid at all times only for short-wavelength vortical perturbations
($kH \gg 1$).} The amplitude of a non-vortical (compressive) shwave
satisfies the differential equation
\begin{equation}\label{SOUND}
\dv{^2 \delta v_{yc}}{t^2} + \left(c_s^2 k^2 + \Omega^2\right) \delta v_{yc} = 0,
\end{equation}
the solutions of which are parabolic cylinder functions. See \cite{jg05}
for further details on the shwave solutions.

\begin{figure}
\plottwo{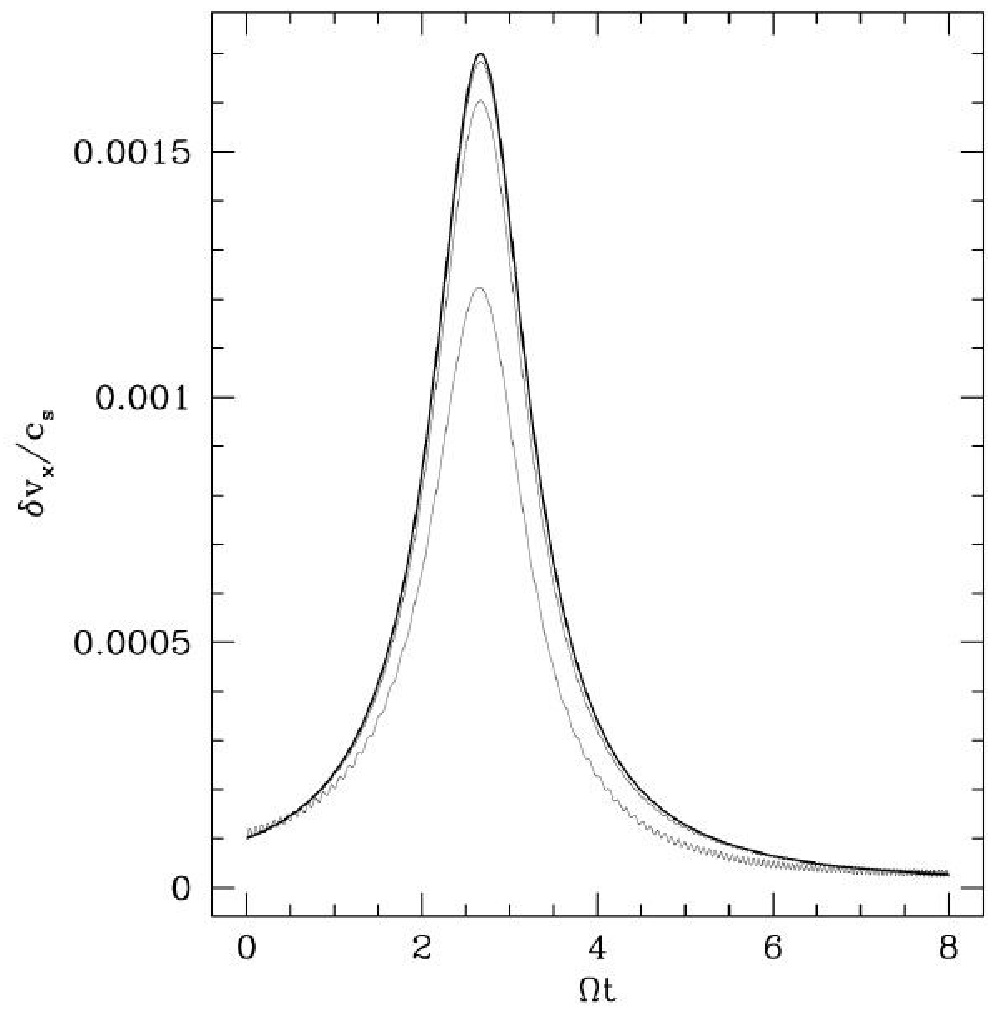}{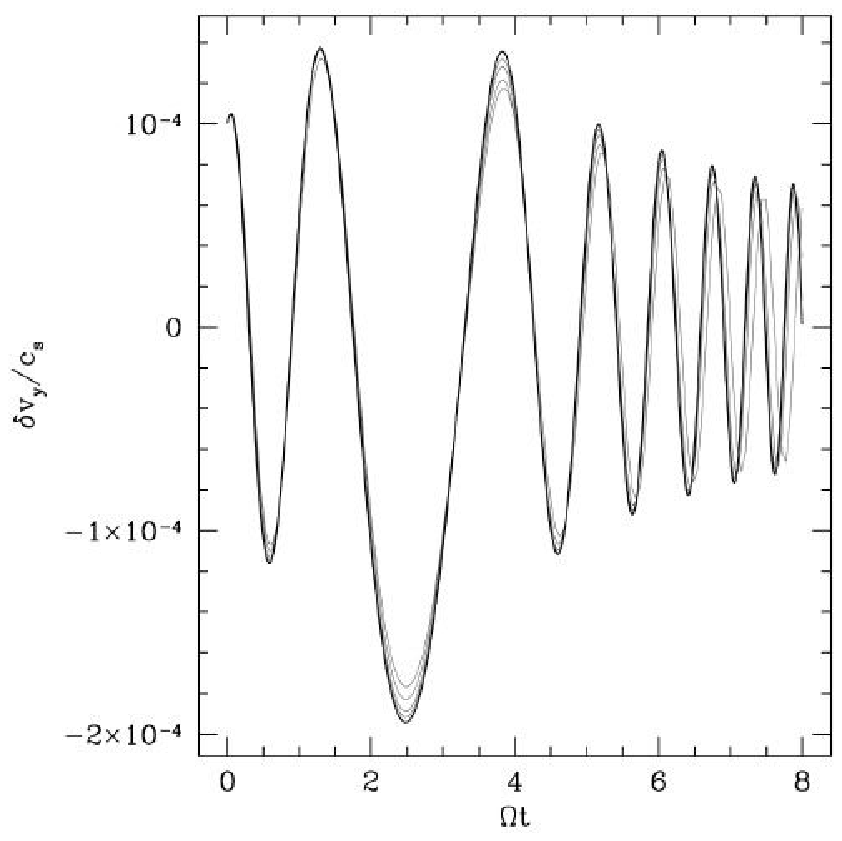}
\caption{
Evolution of velocity amplitudes for a vortical (left) and non-vortical
(right) shwave. The heavy line is the analytic result, and the light lines
are numerical results with (in order of increasing accuracy) $N_x = N_y
= 32, 64, 128$ and $256$.
}
\label{pap3f1}
\end{figure}

Figure~\ref{pap3f1} compares the numerical evolution of both vortical and
compressive shwave amplitudes with their analytic solutions.  The initial
shwave vector ($k_{x0}, k_y$) is ($-16\pi/L_x, 4\pi/L_y$) for the
vortical shwave and ($-8\pi/L_x, 2\pi/L_y$) for the compressive shwave.
The other model parameters are the same as those in the fiducial run, except
that $L = 0.5H$ for the vortical-shwave evolution since $k_y H \gg 1$ is
required to prevent mixing between vortical and non-vortical shwaves
near $\tau = 0$.  The shwaves are well resolved until the radial
wavelength $\lambda_x = 4 \times \Delta x$, and the code is capable of
tracking both potential-vorticity and compressive perturbations with
high accuracy.

\section{Results}

\begin{figure}
\plotone{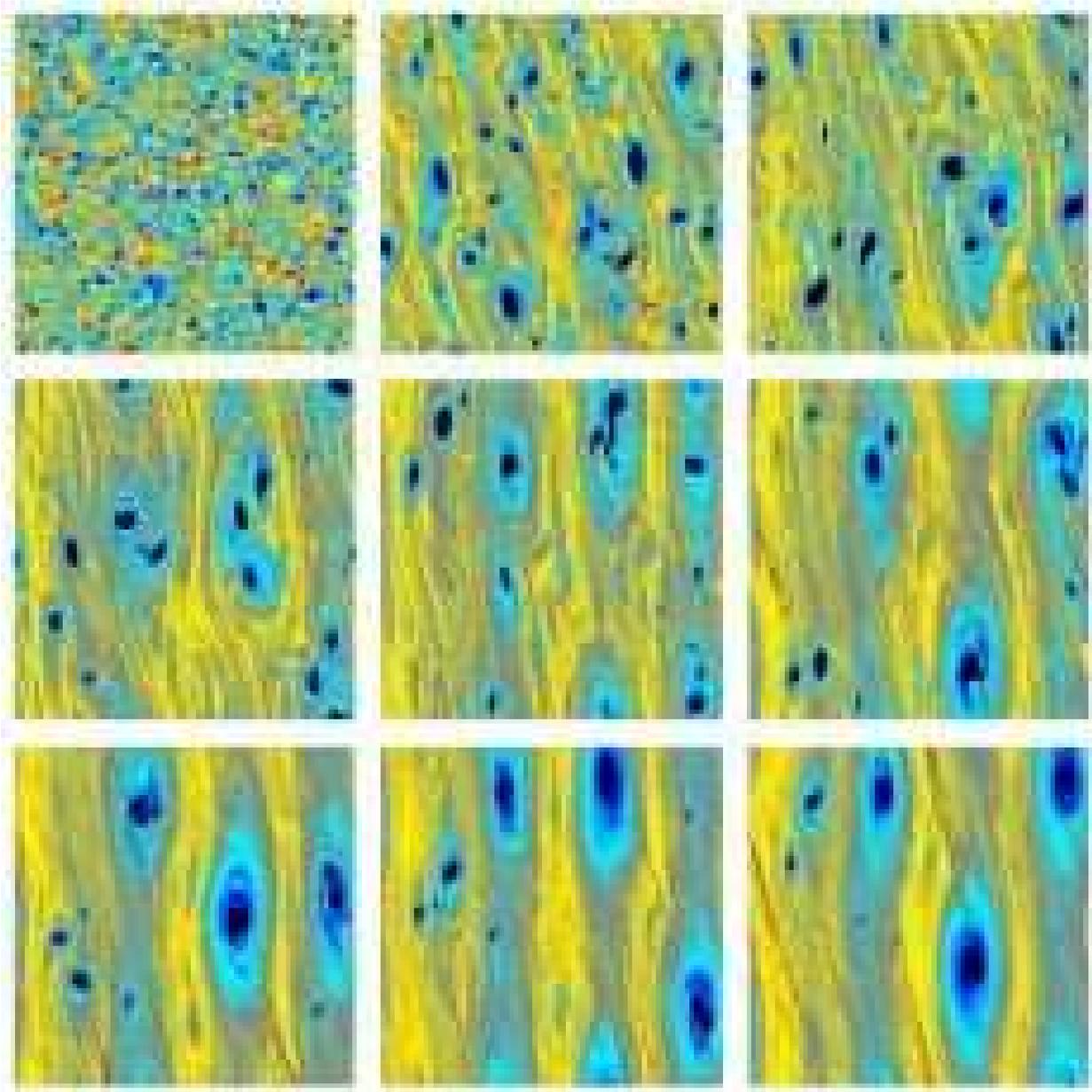}
\caption{
Panels show the evolution of the potential vorticity in the fiducial
run.  The size is $4 H \times 4 H$ and the numerical resolution is
$1024^2$.  The initial conditions are shown in the upper left corner,
and the other frames follow in lexicographic order at intervals of $22.2
\Omega^{-1}$.  Dark shades (blue and black in electronic edition) indicate 
potential vorticity smaller than $\Omega/(2 \Sigma_0)$; light shades 
(yellow and red in electronic edition) indicate positive potential vorticity
perturbations.  Evidently only the ``anticyclonic'' (negative potential
vorticity perturbation) vortices survive.  Each vortex sheds sound
waves, which steepen into trailing shocks.
}
\label{pap3f2}
\end{figure}

The evolution of the potential vorticity in our fiducial run is shown in
Figure~\ref{pap3f2}.  The snapshots are shown in lexicographic order beginning with
the initial conditions, which have equal positive and negative $\delta \xi$.

\begin{figure}
\plottwo{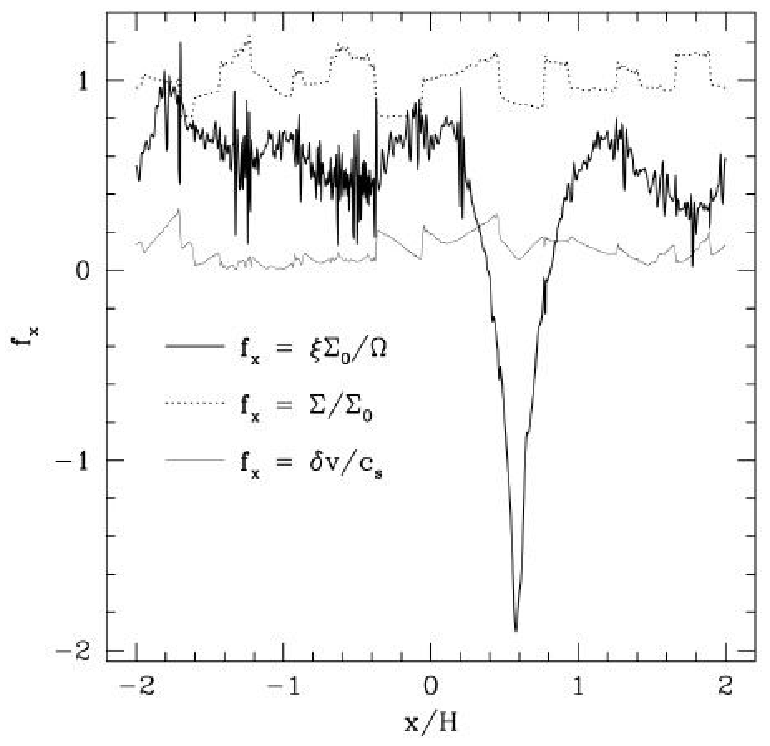}{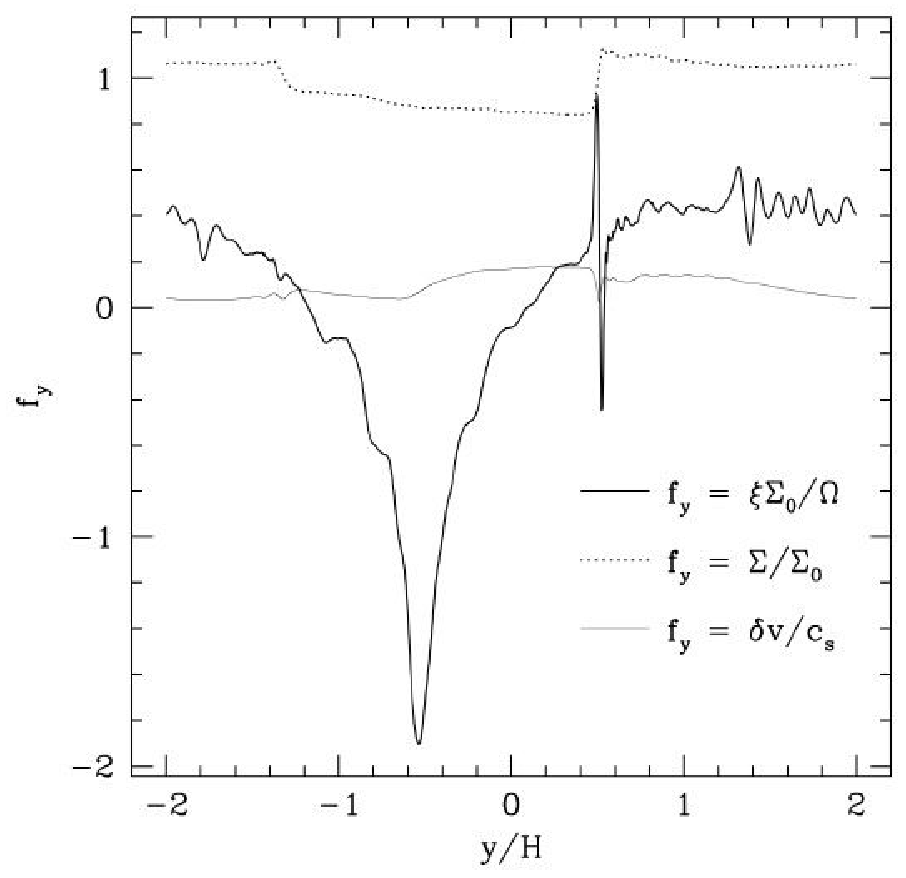}
\caption{
Radial (left) and azimuthal (right) slice of a vortex at the end of the
fiducial run.  The heavy line shows the potential vorticity, the light
line shows the magnitude of the velocity and the dotted line shows the
surface density.
}
\label{pap3f3}
\end{figure}

One of the most remarkable features of the fiducial run evolution is the
appearance of comparatively-stable, long-lived vortices.  These vortices
have negative $\delta \xi$ and are therefore dark in Figure~\ref{pap3f2}.  Similar
vortices have been seen by \cite{gl99,gl00}, \cite{li01} and \cite{ur04}.
Cross sections of one of the vortices at the end of the run are shown in 
Figure~\ref{pap3f3}.  In our models the vortices are not associated with easily 
identifiable features in the surface density, since the perturbed vorticity is 
not large enough to require, through the equilibrium condition, an order unity
increase in the local pressure.

\begin{figure}
\plotone{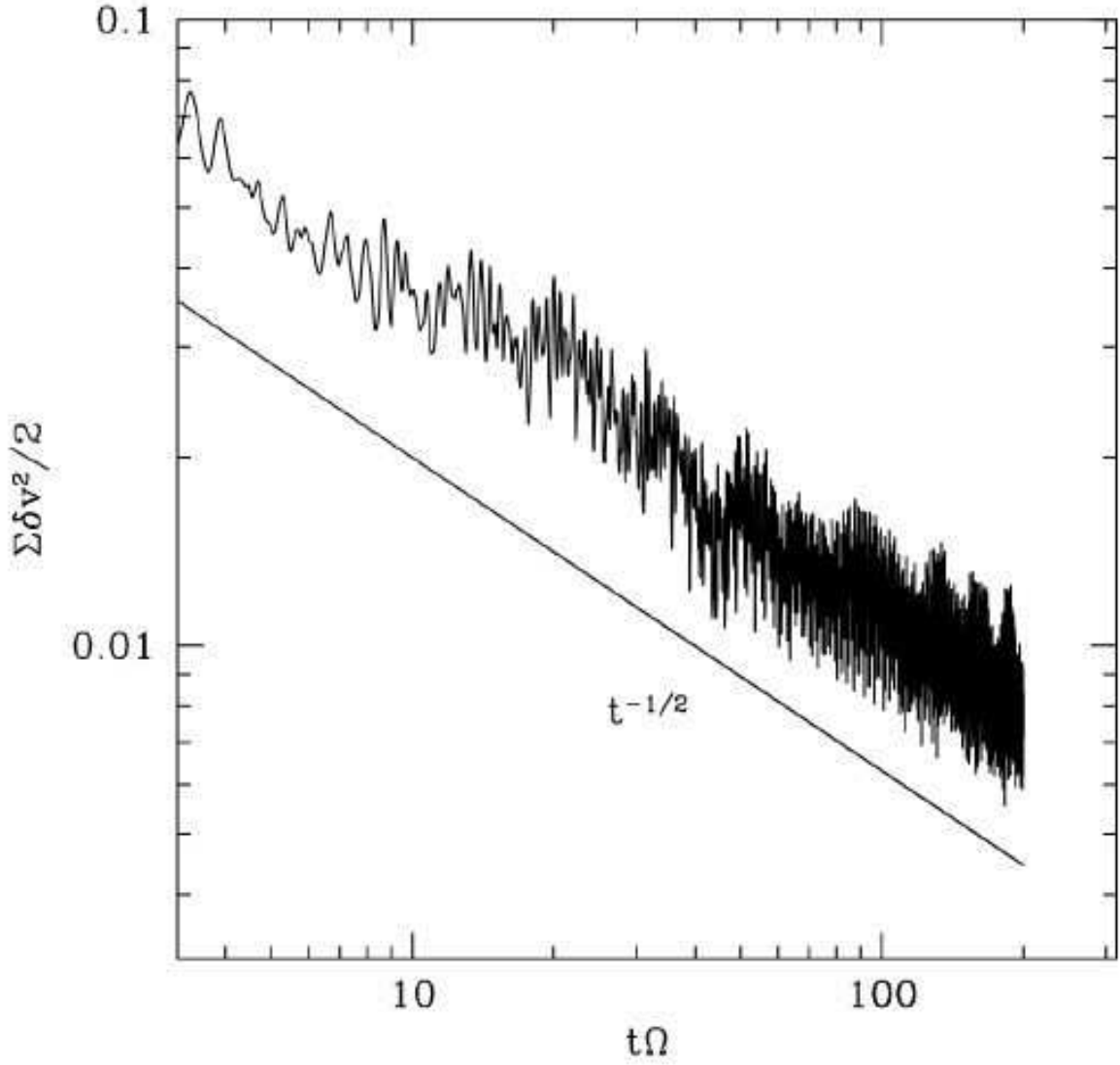}
\caption{
Evolution of kinetic energy in time for the fiducial run, on a log-log
scale.  The solid line shows a $t^{-1/2}$ decay for comparison purposes.
}
\label{pap3f4}
\end{figure}

While the vortices are long-lived, they do decay.  Figure~\ref{pap3f4} shows the
evolution of the perturbed (noncircular) kinetic energy
\begin{equation}
E_K \equiv {1\over{2}} \Sigma (\delta v_x^2 + \delta v_y^2)
\end{equation}
in the fiducial run.  Evidently the kinetic energy decays approximately
as $t^{-1/2}$ (which is remarkable in that, if the vortices would
correspond to features in {\it luminosity} that decay as $t^{-1/2}$,
they could produce flicker noise; see \citealt{press78}).  Runs with half and
twice the resolution decay in the same fashion, but if the resolution is
reduced to $64^2$ the kinetic energy decays exponentially.
Resolution of at least 128 zones per scale height appears to be
required.

What is even more remarkable is that the vortices are associated with an
outward angular momentum flux, due to the driving of compressive
motions by the vortices.  Figure~\ref{pap3f5} shows the evolution 
of the dimensionless angular momentum flux
\begin{equation}
\alpha \equiv {1\over{L_x L_y\Sigma_0 c_s^2}} \int \Sigma \delta v_x \delta v_y dx dy
\end{equation}
for models with a variety of resolutions.  The data has been boxcar
smoothed over an interval $\Delta t = 10 \Omega^{-1}$ to make the plot
readable.  Again, a resolution of at least $512^2$ appears to be
required for a converged measurement of the shear stress.  For the most
highly resolved models $\alpha$ evolves like the kinetic energy,
$\propto t^{-1/2}$.

Compressibility is crucial for development of the angular momentum flux.
We have demonstrated this in two ways.  First, we have taken the
fiducial run and decomposed the velocity field into a compressive and an
incompressive part (i.e., into potential and solenoidal pieces in Fourier
space) and measured the stress associated with each.  For a
set of snapshots taken from the last half of the fiducial run, the
average total $\alpha = 0.0036$; the incompressive component is
$\alpha_i = -0.0006$; the compressive component is $\alpha_c = 0.0032$.
The remaining alpha $\alpha_x = 0.00099$ is in cross-correlations
between the incompressive and compressive pieces of the velocity field.
As argued in \cite{bal00} and \cite{bal03}, both incompressive trailing
shwaves and incompressible turbulence transport angular
momentum inward, whereas trailing compressive disturbances transport
angular momentum outward. Our negative (positive) value for $\alpha_i$
($\alpha_c$) is consistent with this.

Second, we have reduced the size of the model and reduced the amplitude
of the initial perturbation so that it scales with the shear velocity at
the edge of the model (constant ``intensity'' of the turbulence, in
Umurhan and Regev's parlance).  Thus the Mach number of the turbulence
is reduced in proportion to the size of the box.  We have compared four
models, with $L = (4, 2, 1, 0.5) H$ and $\sigma = (0.8, 0.4, 0.2, 0.1)
c_s$.  We would expect the lower Mach number models to have 
smaller-amplitude compressive velocity fields and therefore, consistent 
with the above results, smaller angular momentum flux $\alpha$.  
Averaging over the second half of the simulation, we find $\alpha =
(0.0031, 0.0018, 7.2 \times 10^{-5}, -9.5 \times 10^{-7})$.

\begin{figure}
\plotone{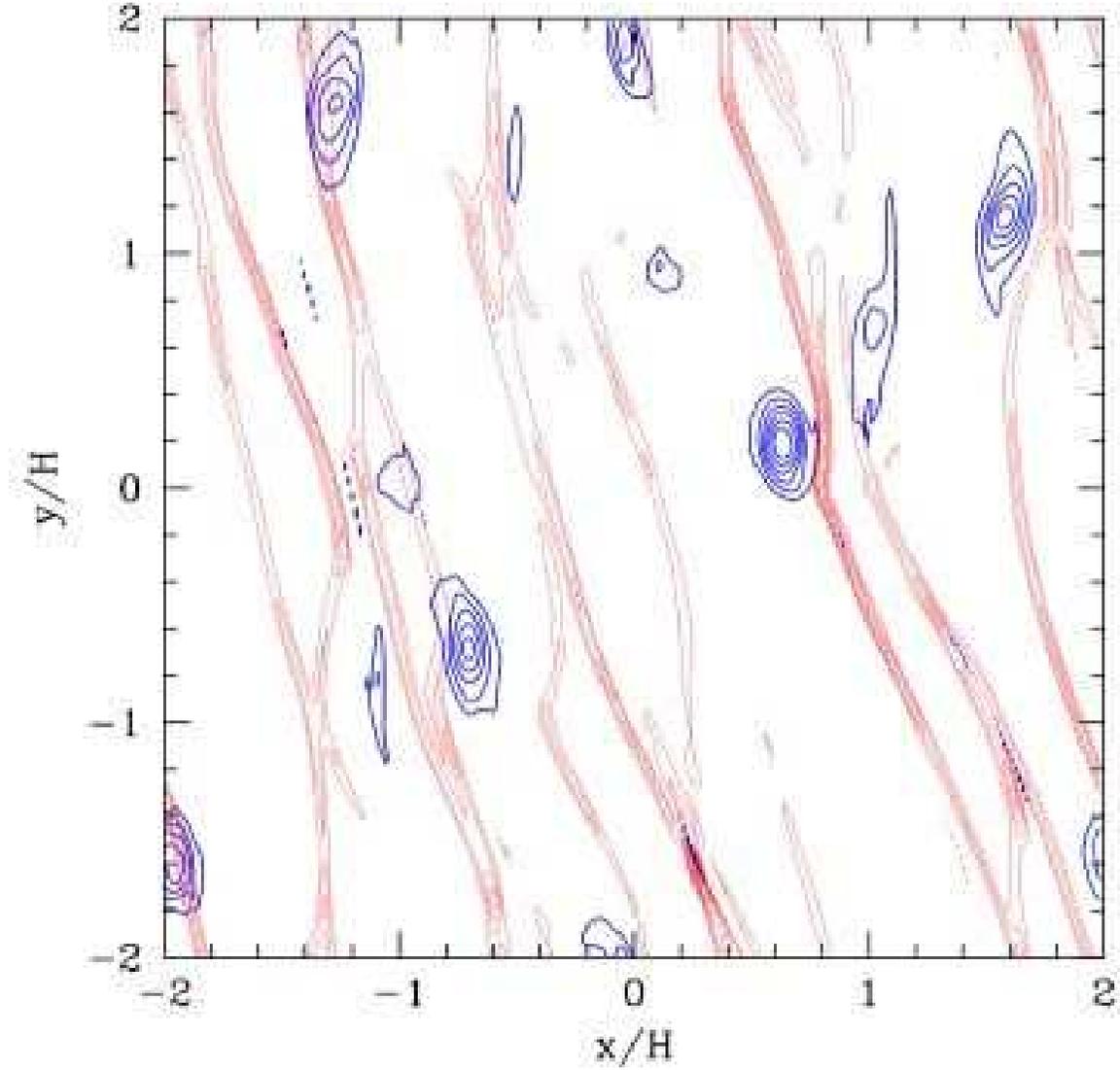}
\caption{
Snapshot of the velocity divergence superimposed on the potential
vorticity in a medium-resolution ($256^2$) version of the fiducial run.
The thin (red in electronic edition) contours indicate negative divergence 
and are associated with shocks. The thick (blue in electronic edition)
contours indicate negative potential vorticity.
}
\label{pap3f6}
\end{figure}

An additional confirmation of our overall picture can be seen in
Figure~\ref{pap3f6}, in which we show a snapshot of the velocity divergence 
superimposed on the potential vorticity for a medium-resolution
($256^2$) version of the fiducial run.\footnote{At higher resolutions, 
shocks are generated earlier in the simulation from smaller vortices, 
and it is more difficult to see the effect we are describing due to the 
random nature of the vortices at this early stage.} The position of
the shocks with respect to the vortices in this figure is consistent 
with our interpretation that the former are generated by the latter. 

The smallest of our simulations ($L = 0.5H$) is nearly incompressible,
but we continue to observe $t^{-1/2}$ decay (least squares fit power law
is $-0.49$) at late times.  The reason that we see decay while
\cite{ur04} do not may be that: (1) the remaining compressibility in our
model causes added dissipation; (2) the numerical dissipation in our
code is larger than that of \cite{ur04}; (3) the code used by
\cite{ur04} could somehow be aliasing power from trailing shwaves to
leading shwaves (although they do explicitly discuss, and dismiss, this
possibility).

\begin{figure}
\plotone{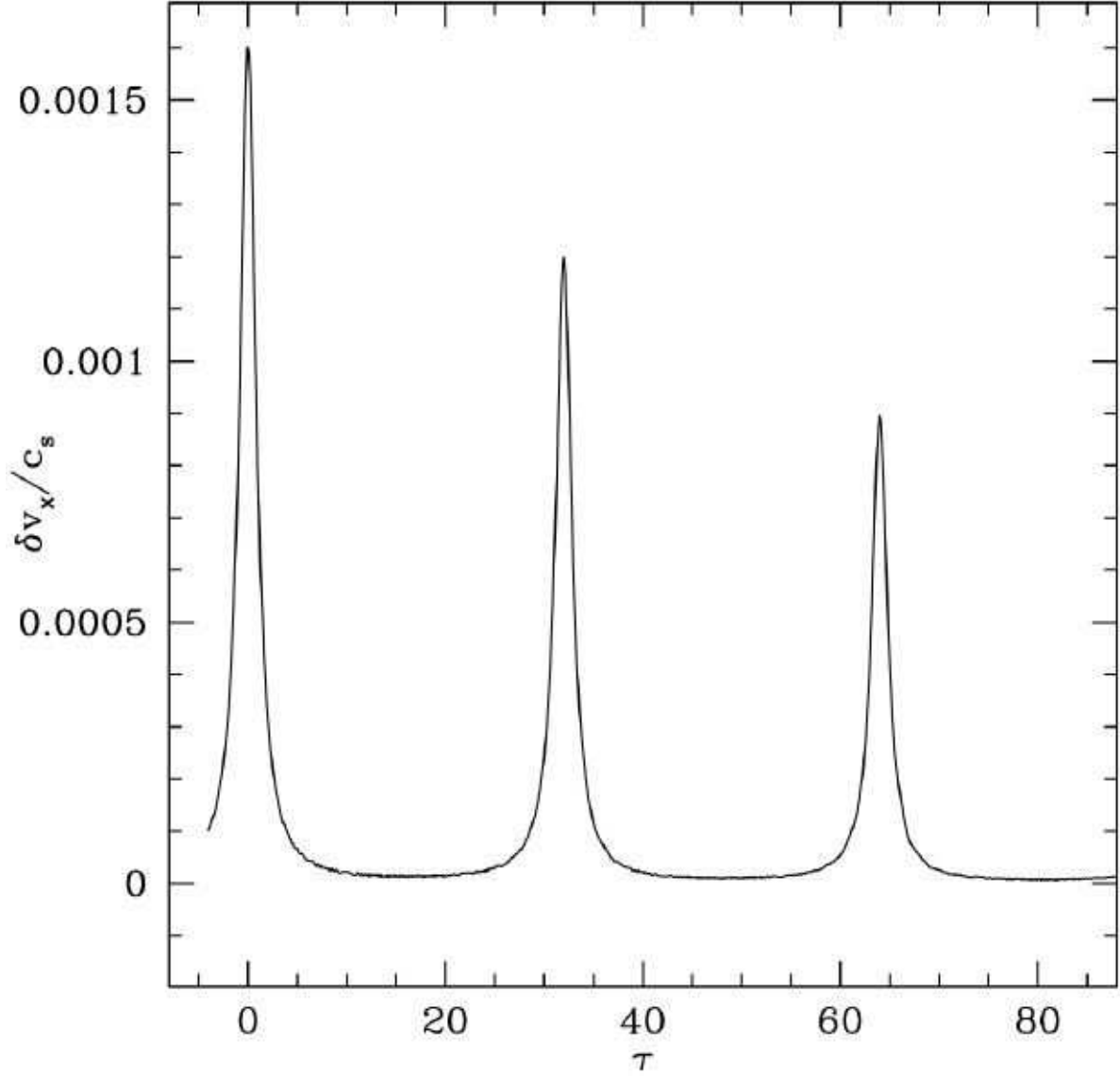}
\caption{
Evolution of a vortical shwave amplitude in a low-resolution ($64^2$)
run, in units of $\tau$. The initial shwave vector ($k_{x0},k_y$) is
($-16\pi/L_x, 4\pi/L_y$), corresponding to $\tau_0 = -4$. The interval
between successive peaks (a numerical effect due to aliasing) is $\tau =
N_x/n_y$, where $n_y = 2$ is the azimuthal wavenumber.
}
\label{pap3f7}
\end{figure}

To highlight the dangers of aliasing for our finite-difference code, in
Figure~\ref{pap3f7} we show the evolution of a vortical shwave amplitude at low
resolution ($64^2$), in units of $\tau$. We use the same parameters as  
those in our linear-theory test (Figure~\ref{pap3f1}), for which the initial shwave 
vector corresponds to $\tau_0 = -4$. The initially-leading shwave swings 
into a trailing shwave, the radial wavelength is eventually lost near the
grid scale, and due to aliasing the code picks up the evolution of the 
shwave again as a leading shwave.  Repeating this test at higher 
resolutions indicates that successive swings from leading to trailing
occur at an interval of $\tau = N_x/n_y$, where $n_y = 2$ is the 
azimuthal wavenumber of the shwave.  This is equivalent to $k_x(t) = 
2\pi/\Delta x$. The decay of the successive linear solutions with time 
is due to numerical diffusion.  

Figure~\ref{pap3f7} suggests that it is easier to inject power into the simulation
due to aliasing rather than to remove power due to numerical diffusion.
We do not believe, however, that aliasing is affecting our
high-resolution results.  In addition, if we assume that the flow in our
simulations can be modeled as two-dimensional Kolmogorov turbulence,
then $\delta v_{rms} \sim \lambda^ {1/3}$, where $\delta v_{rms}$ is 
the rms velocity variation across a scale $\lambda$.  The velocity due to the
mean shear at these scales is $\delta v_{shear} \sim q \Omega \lambda$,
and $\delta v_{rms}/\delta v_{shear} \sim \lambda^{-2/3}$. The velocities at
the smallest scales are thus dominated by turbulence rather than by the
mean shear. This conclusion is supported by the convergence of our
numerical results at high resolution.

\begin{figure}
\plotone{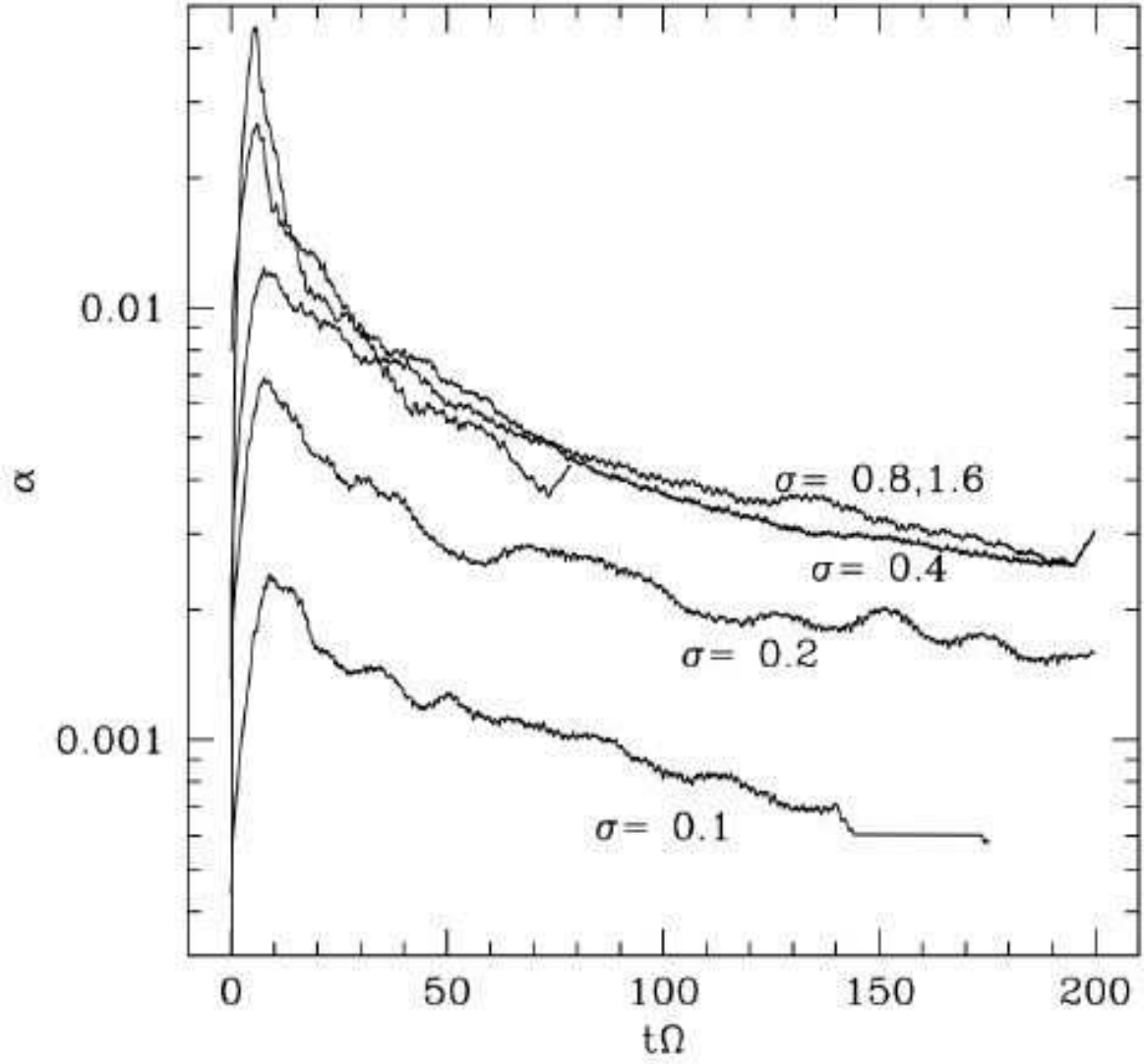}
\caption{
Evolution of the shear stress $\alpha$ in a set of runs at with varying
initial $\sigma$.  Apparently for low values of $\sigma$ the shear
stress is reduced, but for initial Mach number near $1$ the stress
saturates.  All runs have $L = 4 H$.
}
\label{pap3f8}
\end{figure}

\begin{figure}
\plotone{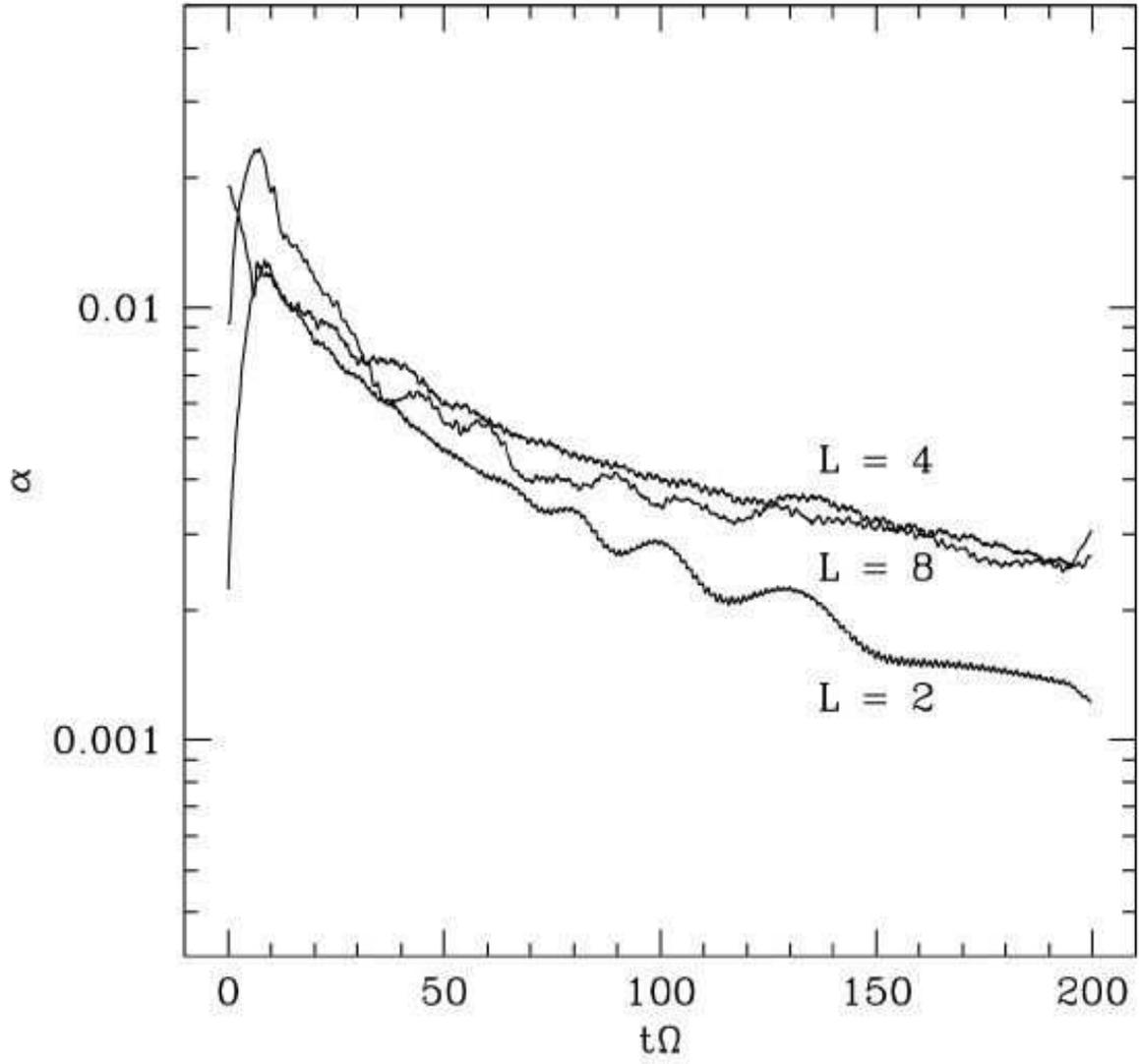}
\caption{
Evolution of the shear stress $\alpha$ in a set of runs at with varying
initial $L$, but the same initial Mach number $\sigma$.
}
\label{pap3f9}
\end{figure}

Our model contains two additional numerical parameters: the size $L$ and
the initial turbulence amplitude $\sigma$. Figure~\ref{pap3f8} shows the evolution
of $\alpha$ for several values of $\sigma$.  Evidently for small enough
values of $\sigma$ the $\alpha$ amplitude is reduced, but for near-sonic
initial Mach numbers the $\alpha$ amplitude saturates (or at least the
dependence on $\sigma$ is greatly weakened).  Figure~\ref{pap3f9} shows the
evolution for several values of $L$ but the same initial $\sigma$ and
the identical initial power spectrum. For large enough $L$ the shear
stress appears to be independent of $L$.  

\begin{figure}
\plottwo{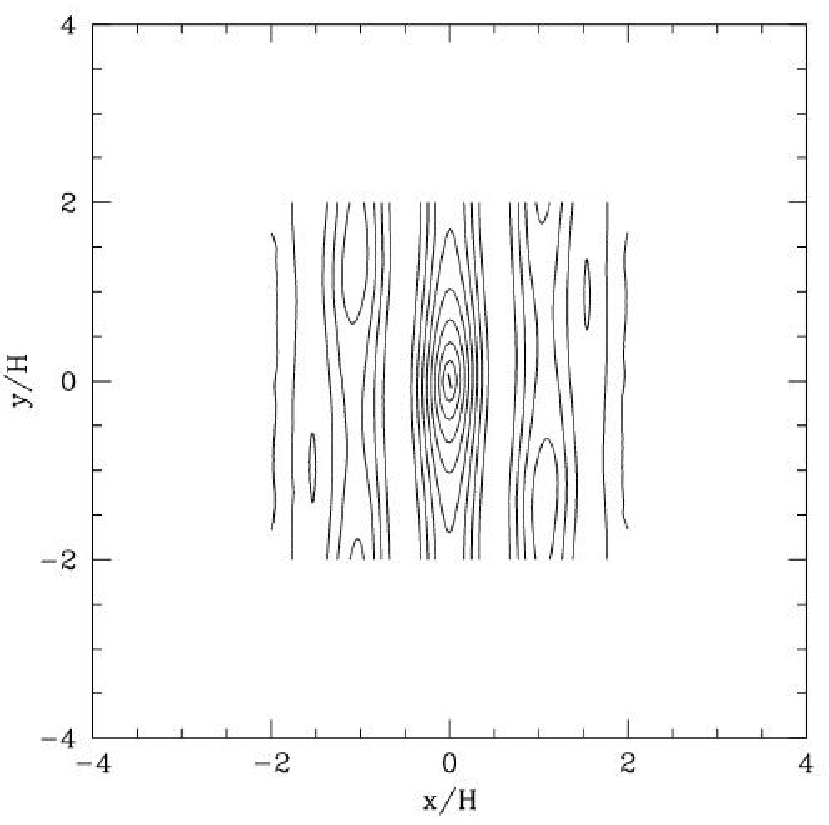}{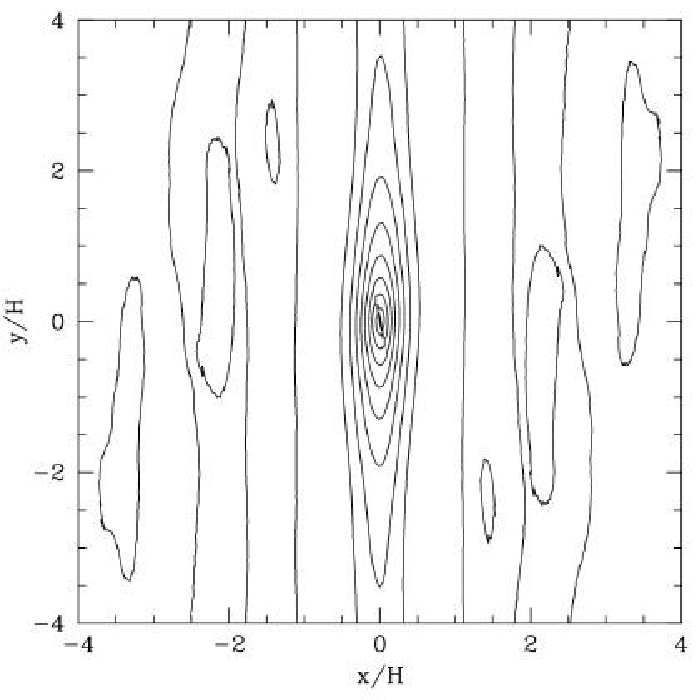}
\caption{
Autocorrelation function of the potential vorticity $\xi$ for the fiducial
model with $L = 4H$ (left) and a model with $L = 8H$ (right).
}
\label{pap3f10}
\end{figure}

Finally, we have studied the autocorrelation function of the potential
vorticity as a means of characterizing structure inside the flow.
Figure~\ref{pap3f10} shows the autocorrelation function measured in the fiducial
model and in an otherwise identical model with $L = 8H$.  Evidently 
the potential vorticity is correlated over about one-half a scale height
in radius, independent of the size of the model.  This supports the idea
that compressive effects limit the size of the vortices, since the shear
flow becomes supersonic across a vortex of size $\sim H$ \citep{bs95,li01}.

\section{Conclusion}

The presence of long-lived vortices in weakly-ionized disks may be an
integral part of the angular momentum transport mechanism in these
systems. The key result we have shown here is that compressibility of
the flow is an extremely important factor in providing a significant,
positively-correlated average shear stress with its associated outward
transport of angular momentum. Previous results using a local model
have assumed incompressible flow and either report no angular momentum
transport \citep{ur04} or report a value ($\alpha \sim 10^{-5}$, \citealt{
bm05}) that is two orders of magnitude lower than what we find when we
include the effects of compressibility. Global simulations
\citep{gl99,gl00,li01} have a difficult time accessing the high
resolution that we have shown is required for a significant shear
stress due to compressibility.

Our work leaves open the key question of what happens in three
dimensions. Our vortices, which have radial and azimuthal extent
$\lesssim H$, are inherently three-dimensional. Three-dimensional
vortices are susceptible to the elliptical instability \citep{ker02} and
are likely to be destroyed on a dynamical timescale. The fact that
vortices persist in our two-dimensional simulations and not in the
local (three-dimensional) shearing-box calculations of \cite{bh96} is
likely due to dimensionality. The recent numerical results of \cite{bm05}
indicate that vortices near the disk midplane are quickly destroyed,
whereas vortices survive if they are a couple of scale heights away
from the midplane. Strong vertical stratification away from the midplane 
may enforce two-dimensional flow and allow the vortices that we consider 
here to survive.

The initial conditions in \cite{bm05} are analytic solutions for
two-dimensional vortices that are stacked into a three-dimensional
column. The stable, off-midplane vortices apparently arise due to the
breaking of internal gravity waves generated by the midplane vortices
before they become unstable. There is also an unidentified instability
that breaks a single off-midplane vortex into several vortices.  These
simulations leave open the question of whether stable off-midplane
vortices can be generated from a random set of initial vorticity
perturbations rather than the special vortex solutions that are imposed.

\begin{figure}
\plotone{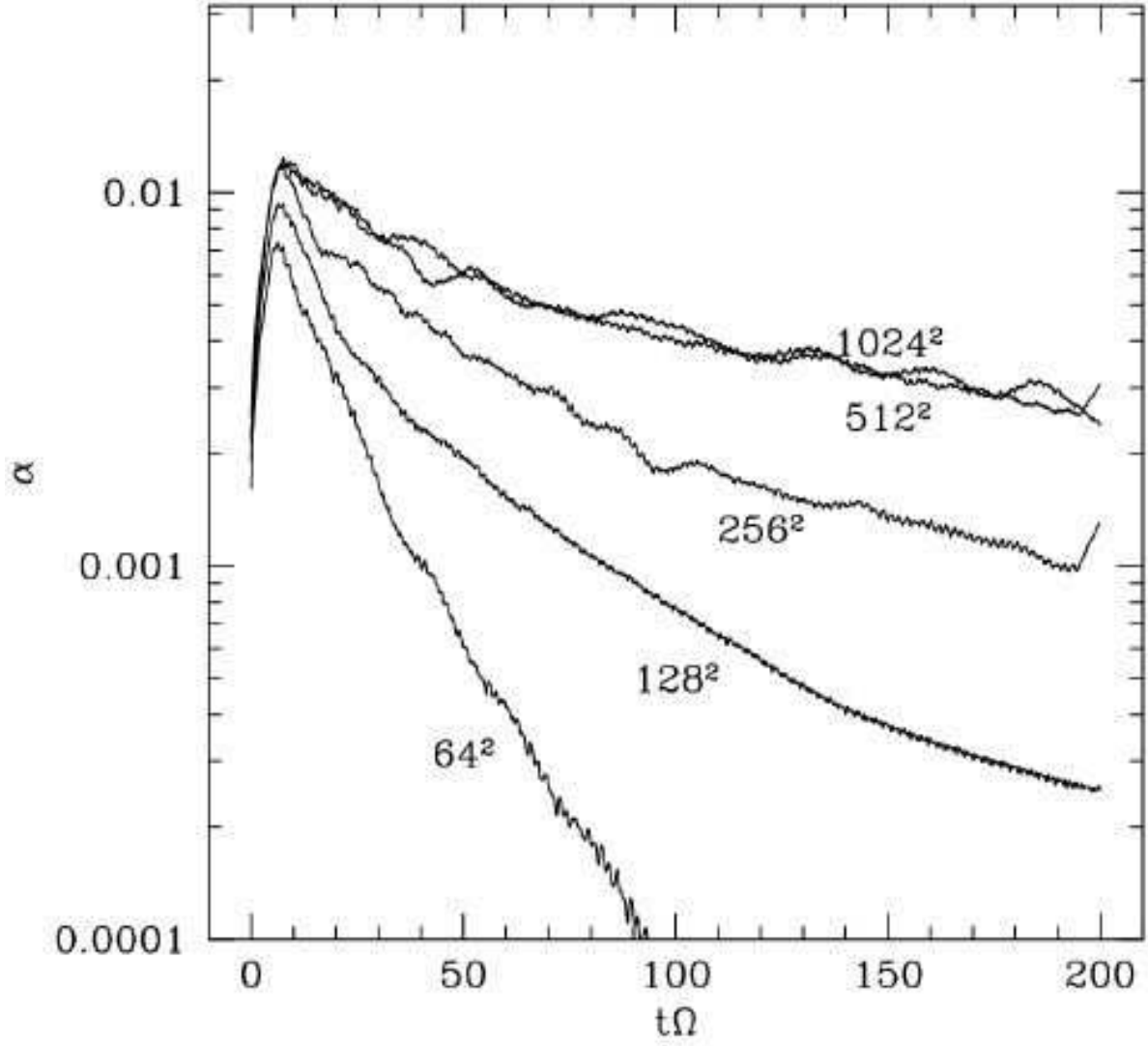}
\caption{
Evolution of the shear stress $\alpha$ in the fiducial run and a set of
runs at lower resolutions.
}
\label{pap3f5}
\end{figure}

Our work also leaves open the key question of what generates the initial
vorticity. One possibility is that material builds up at particular
radii in the disk, resulting in a global instability (e.g.
\citealt{pp84,pp85}) and a breakdown of the flow into vortices
\citep{li01}. A second possibility for vortex generation in variable
systems is that the MHD turbulence, which likely operates during an
outburst but decays as the disk cools \citep{gm98}, leaves behind some
residual vorticity. The viability of such a mechanism could be tested
with non-ideal MHD simulations such as those of \cite{fs03} and
\cite{ss03}. A third possibility is that differential illumination of the disk
somehow produces vorticity.  Since the temperature of most circumstellar
disks is controlled by stellar illumination, small variations in
illumination could produce hot and cold spots in the disk that interact
to produce vortices. A fourth possibility is the generation of vorticity
via baroclinic instability, which is likely to operate in disks
whose vertical stratification is close to adiabatic \citep{ks86}. The
nonlinear outcome of this instability in planetary atmospheres is the
formation of vortices, although it is far from clear that the same outcome
will occur in disks. Finally, we note that a residual amount of vorticity
can be generated from finite-amplitude compressive perturbations. We have
performed a series of runs with zero initial vorticity and perturbation
wavelengths on the order of the scale height, and the results are
qualitatively similar to Figure~\ref{pap3f5} with the shear stress
reduced by nearly two orders of magnitude.

\acknowledgments

This work was supported by NSF grant AST 00-03091 and PHY 02-05155,
and a Drickamer Fellowship for BMJ.

\newpage

\newpage

\begin{thebibliography}{}

\bibitem[Balbus(2000)]{bal00} Balbus, S.~A.\ 2000, \apj, 534, 420 

\bibitem[Balbus(2003)]{bal03} Balbus, S.~A.\ 2003, \araa, 41, 555

\bibitem[Balbus \& Hawley(1991)]{bh91} Balbus, S.~A.~\&
	Hawley, J.~F.\ 1991, \apj, 376, 214

\bibitem[Balbus et al.(1996)]{bh96} Balbus, S.~A., Hawley,
	J.~F., \& Stone, J.~M.\ 1996, \apj, 467, 76

\bibitem[Balbus \& Hawley(1998)]{bh98} Balbus, S.~A., \& 
	Hawley, J.~F.\ 1998, Reviews of Modern Physics, 70, 1

\bibitem[Baptista et al.(1998)]{bap98} Baptista, R., Horne, 
	K., Wade, R.~A., Hubeny, I., Long, K.~S., \& 
	Rutten, R.~G.~M.\ 1998, \mnras, 298, 1079

\bibitem[Baptista(2004)]{bap04} Baptista, R.\ 2004, 
	Astronomische Nachrichten, 325, 181 

\bibitem[Barge \& Sommeria(1995)]{bs95} Barge, P., \&
	Sommeria, J.\ 1995, \aap, 295, L1

\bibitem[Barranco \& Marcus(2005)]{bm05} Barranco, J.~A., \& 
	Marcus, P.~S.\ 2005, ArXiv Astrophysics e-prints, astro-ph/0501267 

\bibitem[Blaes \& Hawley(1988)]{bh88} Blaes, O.~M., \& 
	Hawley, J.~F.\ 1988, \apj, 326, 277

\bibitem[Desch(2004)]{des04} Desch, S.~J.\ 2004, \apj, 608, 509

\bibitem[Fleming \& Stone(2003)]{fs03} Fleming, T., \& 
	Stone, J.~M.\ 2003, \apj, 585, 908 

\bibitem[Fromang et al.(2002)]{ftb02} Fromang, S., Terquem, 
	C., \& Balbus, S.~A.\ 2002, \mnras, 329, 18 

\bibitem[Gammie \& Menou(1998)]{gm98} Gammie, C.~F.~\& 
	Menou, K.\ 1998, \apjl, 492, L75

\bibitem[Gammie(2001)]{gam01} Gammie, C.~F. 2001, \apj, 553, 174

\bibitem[Godon \& Livio(1999)]{gl99} Godon, P., \& 
	Livio, M.\ 1999, \apj, 523, 350

\bibitem[Godon \& Livio(2000)]{gl00} Godon, P., \& 
	Livio, M.\ 2000, \apj, 537, 396

\bibitem[Gough(1969)]{gou69} Gough, D.~O.\ 1969, Journal of
Atmospheric Sciences, 26, 448

\bibitem[Hawley(1987)]{haw87} Hawley, J.~F.\ 1987, \mnras, 225, 677

\bibitem[Hawley(1990)]{haw90} Hawley, J.~F.\ 1990, \apj, 356, 580

\bibitem[Hawley, Gammie, \& Balbus(1995)]{hgb95} Hawley, J.~F., 
	Gammie, C.~F., \& Balbus, S.~A.\ 1995, \apj, 440, 742

\bibitem[Johnson \& Gammie(2005)]{jg05} Johnson, B.~M.~\&
	Gammie, C.~F.\ 2005, \apj, submitted

\bibitem[Kerswell(2002)]{ker02} Kerswell, R.~R.\ 2002, Annual 
	Review of Fluid Mechanics, 34, 83 

\bibitem[Klahr \& Bodenheimer(2003)]{kb03} Klahr, H.~H.~\&
	Bodenheimer, P.\ 2003, \apj, 582, 869

\bibitem[Knobloch \& Spruit(1986)]{ks86} Knobloch, E., \& 
	Spruit, H.~C.\ 1986, \aap, 166, 359

\bibitem[Kunz \& Balbus(2004)]{kb04} Kunz, M.~W.~\& 
	Balbus, S.~A.\ 2004, \mnras, 348, 355

\bibitem[Li, Finn, Lovelace, \& Colgate(2000)]{li00} Li, H., Finn, J.~M., 
	Lovelace, R.~V.~E., \& Colgate, S.~A.\ 2000, \apj, 533, 1023 

\bibitem[Li, Colgate, Wendroff, \& Liska(2001)]{li01} Li, H., Colgate, 
	S.~A., Wendroff, B., \& Liska, R.\ 2001, \apj, 551, 874 

\bibitem[Lovelace, Li, Colgate, \& Nelson(1999)]{love99} Lovelace, R.~V.~E., 
	Li, H., Colgate, S.~A., \& Nelson, A.~F.\ 1999, \apj, 513, 805

\bibitem[Masset(2000)]{mass00} Masset, F.\ 2000, \aaps, 141, 165

\bibitem[Menou(2000)]{men00} Menou, K.\ 2000, Science, 288, 2022

\bibitem[Menou \& Quataert(2001)]{mq01} Menou, K., \& 
	Quataert, E.\ 2001, \apj, 552, 204 

\bibitem[Narayan et al.(1987)]{ngg87} Narayan, R., Goldreich, 
	P., \& Goodman, J.\ 1987, \mnras, 228, 1 

\bibitem[Papaloizou \& Pringle(1984)]{pp84} Papaloizou, 
	J.~C.~B., \& Pringle, J.~E.\ 1984, \mnras, 208, 721 

\bibitem[Papaloizou \& Pringle(1985)]{pp85} Papaloizou, 
	J.~C.~B., \& Pringle, J.~E.\ 1985, \mnras, 213, 799 

\bibitem[Press(1978)]{press78} Press, W.~H.\ 1978, Comments on 
	Astrophysics, 7, 103 

\bibitem[Sano \& Stone(2003)]{ss03} Sano, T., \& Stone, 
	J.~M.\ 2003, \apj, 586, 1297 

\bibitem[Stone, Gammie, Balbus, \& Hawley(2000)]{sgbh00} 
	Stone, J.~M., Gammie, C.~F., Balbus, S.~A., \& 
	Hawley, J.~F.\ 2000, Protostars and Planets IV, 589

\bibitem[Stone \& Norman(1992)]{sn92} Stone, J.~M.~\& Norman, 
	M.~L.\ 1992, \apjs, 80, 753

\bibitem[Umurhan \& Regev(2004)]{ur04} Umurhan, O.~M.~\& 
	Regev, O.\ 2004, \aap, 427, 855

\end{thebibliography}
\end{document}